# Polarization attractors driven by vector soliton rain

**SERGEY V. SERGEYEV,[1,*] MAHMOUD ELIWA,[1] AND HANI KBASHI[1]**

[1]*Aston Institute of Photonic Technologies, Aston University, Birmingham, B4 7ET, UK*
*Corresponding author: s.sergeyev@aston.ac.uk*

**Abstract:** Soliton rain is a bunch of small soliton pulses slowly drifting nearby the main pulse having the period of a round trip. For Er-doped fiber laser mode-locked by carbon nanotubes, for the first time, we demonstrate both experimentally and theoretically a new type of polarization attractors controllable by the vector soliton rain. With adjusting the pump power, vector soliton rain takes the form of pulses with rotating states of polarization which enable transforming slowly evolving trajectories on the Poincaré sphere from the double-scroll spiral to the circle. The obtained results on controlling complex multisoliton dynamics can be of interest in laser physics and engineering with potential applications in spectroscopy, metrology, and biomedical diagnostics.

## 1. INTRODUCTION

Dissipative solitons (DSs) as localized waves arising from a balance of dissipative and dispersive effects in the open systems is a fundamental paradigm in different areas of science and technology, including nonlinear science, fluid dynamics, neuroscience, telecommunication, and photonics [1-10]. Mode-locked laser (MLL) as a testbed for generating the DSs with controllable parameters is widely used to study short-range and long-range interactions among DSs in the context of unlocking the routes to control the formation of self-stabilized structures in a wide range of distributed systems [3-19].

Short-range interaction through the overlapping of solitons tails or soliton-dispersive wave interaction results in soliton bound states formation with the spacing of the few pulse widths [14-16]. Unlike this, long-range interactions can be driven, for example, by Casimir-like [17, 18], optoacoustic [11-13], and polarization effects [19] and can lead to soliton supramolecular structures in the form of multi-pulsing, harmonic mode-locking, breather-like, and the soliton rain (SR) [11-13, 18-24]. The absence of effective approaches to controlling long-range soliton interactions leads to harmonically mode-locked laser pulses with uncontrollable repetition rates and high timing jitter, drifting soliton rain, or weakly-bound solitons (see references in [25]). The most effective approach for control is based on tailoring optoacoustic effects and dispersive-wave emission. As a result, it was demonstrated that the self-assembly of the optical solitons into stable supramolecular structures can find potential applications in optical information storage and ultrafast laser dynamic waveforms manipulation [25].

The supramolecule in the form of SR presents a bunch of small soliton pulses located nearby the main pulse and coexisting with a continuous wave (cw) background [18-23]. The soliton rain was studied experimentally and theoretically for fiber lasers with different mode-locking mechanisms based on nonlinear polarization rotation (NPR), nonlinear amplified loop mirror (NALM), the figure of eight cavity, graphene, and single-wall carbon nanotubes (SWCNT) [18-23]. Though the variety of mechanisms of SR evolution in different MLLs has been studied in detail, the effective approaches to control the SR have not been revealed yet. In our previous work [19], we found competition between polarization hole burning induced by SR and hole refilling driven by the pump wave, and the state of polarization (SOP) evolution can drive the SR dynamics towards SR pulses merging or repulsing. However, observed fast polarization dynamics and low temporal resolution of polarimeter were obstacles to revealing complex polarization attractors on the Poincaré sphere and so to conclude about approaches to control SR emergence.

Here, we report for the first time the experimental and theoretical results on the polarization attractors mediated by the soliton rain evolution in the laser cavity. The soliton rain dynamics in the fast time scale of the nanosecond range were analyzed by using a fast photodetector and oscilloscope. The slow vector dynamics in the time scale of microseconds was studied by using an in-line polarimeter. To explain the tunability of polarization attractors from the double-scroll spiral to the circle on the Poincaré sphere, a new theoretical model was developed. In this model, the vector SR was presented in the form of the injected signal with rotating states of polarization, increased amplitude of which leads to the attractor's tunability close to the experimentally observed.

## 2. EXPERIMENTAL SETUP AND METHODS

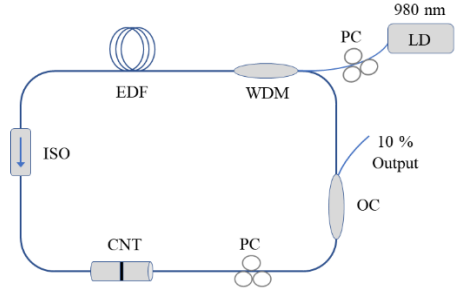

Fig. 1 Erbium-doped fiber laser. EDF: erbium-doped fiber; LD: 980 nm laser diode for the pump; PC: polarization controllers, ISO: optical isolator; WDM: wavelength division multiplexer (WDM), OC: 90:10 output coupler.

The schematic of the unidirectional cavity fiber laser mode-locked is shown in Fig. 1. It has a 14 m long ring cavity corresponding to a round trip time of 70 ns and so the fundamental frequency of mode-locked oscillations is 14.28 MHz. The cavity consists of 13 m standard single-mode optical fiber (SMF-28) with $\beta_2 = -22\ ps^2 km^{-1}$ and a 1 m long erbium-doped fiber (EDF: Liekki Er80-8/125) with the absorption of 80 $dBm^{-1}$ at 1530 nm acting as the gain medium. The EDF is pumped with a 980 nm laser diode (LD) followed by a polarization controller (PC) to adjust the pump light ellipticity. A 980/1550 nm wavelength-division multiplexer (WDM) is used to couple the pump light into the laser cavity. A polarization-independent optical isolator with attenuation of 51 $dB$ is incorporated into the cavity to force the unidirectional operation of the laser. A carbon nanotube (CNT) is used as a saturable absorber to initiate the mode-locking regime, followed by a second PC to tune the linear and circular cavity birefringence. A 90/10 fiber coupler is used to feedback 90% to the laser cavity, and 10% is extracted as the output signal. Instead of using the general approach of splicing fiber to form the ring cavity, all the components are attached through APC fiber connectors. The signal was detected by using a photodetector with a bandwidth of 17 GHz (InGaAsUDP-15-IR-2 FC) connected to a 2.5 GHz sampling oscilloscope (Tektronix DPO7254). An in-line polarimeter (Thorlabs IPM5300) was used to record the state and degree of polarization (SOP and DOP), respectively.

## 3. EXPERIMENTAL RESULTS

The results on the fast dynamics are shown in Fig.2 (a-f). The results have been obtained without adjustment of polarization controllers and with changing pump current J as follows: J=220 mA (a, d), 190 ma (b, e), and 260 mA (c, f). As follows from Fig. 2, the soliton rain changes shape from the condensed phase (Fig. 2 (a, b)) to the bunch of pulses like a burst of soliton rain shown in Fig. 2 (c). As follows from Fig. 2 (d-f), transition to the burst is accompanied by the emergence of the cw component in Fig. 2 (f).

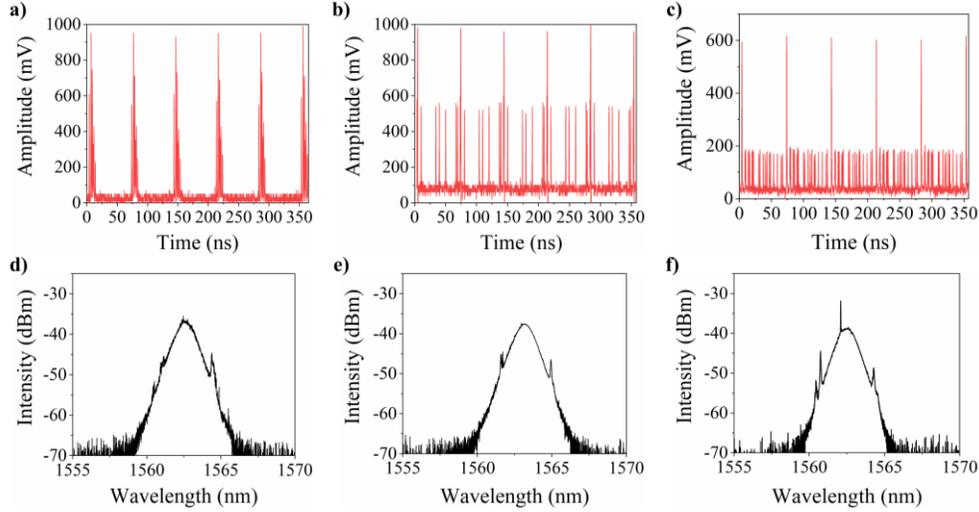

Fig. 2 Fast dynamics. Oscillograms (a-c), and corresponding optical spectra (d-f). Parameters: Pump current J=220 mA (a, d); 190 mA (b, e); 260 mA (c, f).

To study the soliton rain pulses' effect on the state of polarization, we use a polarimeter (resolution of 1 μs and recorded three frames with 1024 samples in each frame) to measure the normalized Stokes parameters $s_1$, $s_2$, $s_3$, and degree of polarization (DOP) which are related to the output powers of two linearly cross-polarized SOPs $I_x$ and $I_y$ and phase difference between them $\Delta\phi$ as follows:

$$S_0 = I_x + I_y, \; S_1 = I_x - I_y, \; S_2 = 2\sqrt{I_x I_y} \cos \Delta\phi, \; S_3 = 2\sqrt{I_x I_y} \sin \Delta\phi,$$
$$s_i = S_i / \sqrt{S_1^2 + S_2^2 + S_3^2}, \quad DOP = \sqrt{S_1^2 + S_2^2 + S_3^2}/S_0, (i = 1,2,3) \quad (1)$$

The obtained results are shown in Fig. 3 (a-i). The slow dynamics corresponds to the cases of the dynamics shown in Fig. 2, i.e., Fig. 3 (a, d, g) – to Fig. 2 (a, d), Fig. 3 (b, e, h) – to Fig. 2 (b, e), and Fig. 3 (c, f, i) – to Fig. 2 (c, f). For the case of the soliton rain condensed phase (Fig. 2 (a, d)), polarization dynamics is shown in Fig. 3 (a, d, g). As follows from Fig.3 (a, b) and our previous publications, spiral attractors can emerge when the in-cavity birefringence is suppressed and the pump wave ellipticity is close to one [26-28]. The specific feature of the spiral attractor is antiphase dynamics of the orthogonal linearly polarized SOPs x and y shown accompanied by switching in Fig. 3 (d) [26-28]. As shown in Fig. 3 (g), the phase difference demonstrates the oscillations and phase slips between states of π/2 and 3π/2. The high DOP of about 90% indicates that polarization evolution is relatively slow and can be mapped with 1 μs resolution (Fig. 3 (g)). If satellite pulses are located close to the main pulse (Fig. 2 (b, e)), the polarization dynamics (Fig. 3 (b, e, h)) looks quite similar to the dynamics shown in Fig. 3 (a, d, g). The main difference is in the oscillatory switching between x and y SOPs (Fig. 3 (e) leading to the slight modification of the spiral attractor in Fig. 3 (b) and oscillatory type of the phase slips in Fig. 3 (h). The dropped DOP to 80% indicates that dynamics is faster than 1 μs. With the increased number of satellite pulses (Fig. 2 (c, f)), polarization attractor changes shape from spiral to circle, as shown in Fig. 3 (c). The dynamics of x and y SOP takes the antiphase form shown in Fig. 3 (f). The phase difference dynamics demonstrates oscillatory behavior between states of π/2 and 3π/2 (Fig. 3 (i)). The high DOP of 95% is an indication of slow dynamics that can be mapped with a resolution of 1 μs (Fig. 3 (i)).

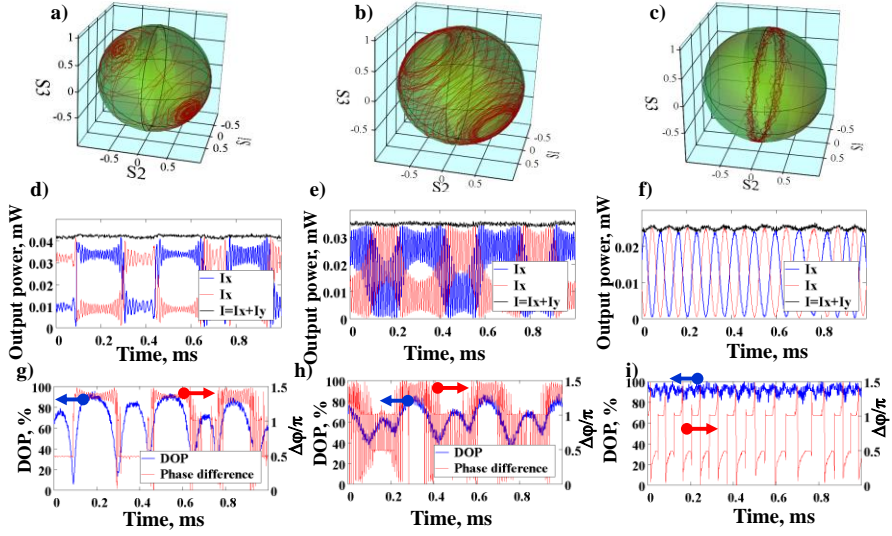

Fig. 3 Slow polarization dynamics: a)- c) trajectories on the Poincaré sphere.); d) - f) The output power vs time for two linearly cross-polarized SOPs $I_x$ (blue line) and $I_y$ (red) and total power $I=I_x+I_y$ (black); g-f) DOP (blue) and the phase difference (red) vs time. Parameters: Pump current J=220 mA (a, d, g); 190 ma (b, e, h); 260 mA (c, f, i).

## 4. THEORETICAL RESULTS

To understand the mechanism of polarization attractor evolution caused by drifting soliton rain, we developed a new vector model of EDFLs from the vector theory developed by Sergeyev and co-workers [11, 26-29]. The model accounts for the linear and circular birefringence and evolution of the laser SOPs and the first excited level population in $Er^{3+}$ doped active medium. The previous model was complemented by the model of the vector soliton rain in the form of an injected signal with periodically evolving orthogonal states of polarization:

$$E_x = a \cdot \cos(\Omega t + \phi_0), E_y = a \cdot \sin(\Omega t + \phi_0) \cdot exp(\Delta\varphi). \qquad (2)$$

Here $a$ is the amplitude of the soliton rain, $\Omega$ is the frequency of oscillations, and $\phi_0$ is the initial phase, $\Delta\varphi$ is the phase difference between the orthogonal SOPs. The model is inspired by the idea that the main pulse produces the polarization hole burning in the orientation distribution of inversion and so SR pulses have SOPs different from the main pulse's SOP [19]. Periodically evolving SOP of the main pulse leads to the oscillation of the SR SOP as shown in Eq. 2. With taking into account Eq. (2), the model found in [11, 26-29] can be presented as follows:

$$\frac{du}{dt} = i\beta u + i\frac{\gamma}{2}\left(|u|^2 u + \frac{2}{3}|v|^2 u + \frac{1}{3}v^2 u^*\right) + D_{xx}u + D_{xy}v + E_x,$$

$$\frac{dv}{dt} = -i\beta v + i\frac{\gamma}{2}\left(|v|^2 v + \frac{2}{3}|u|^2 v + \frac{1}{3}u^2 v^*\right) + D_{xy}u + D_{yy}v + E_y,$$

$$\frac{dn_0}{dt} = \varepsilon\left[I_p + 2R_{10} - \left(1 + \frac{I_p}{2} + \chi R_{10}\right)n_0 - \chi R_{11}n_{12} - \chi n_{22}R_{12}\right],$$

$$\frac{dn_{12}}{dt} = \varepsilon\left[\frac{(1-\delta^2)}{(1+\delta^2)}\frac{I_p}{2} + R_{11} - \left(1 + \frac{I_p}{2} + \chi R_{10}\right)n_{12} - \left(\frac{(1-\delta^2)}{(1+\delta^2)}\frac{I_p}{2} + \chi R_{11}\right)\frac{n_0}{2}\right],$$

$$\frac{dn_{22}}{dt_s} = \varepsilon\left[R_{12} - \left(1 + \frac{I_p}{2} + \chi R_{10}\right)n_{22} - \chi R_{12}\frac{n_0}{2}\right],$$

$$R_{10} = \frac{1}{(1+\Delta^2)}(|u|^2 + |v|^2), \quad R_{11} = \frac{1}{(1+\Delta^2)}(|u|^2 - |v|^2), R_{12} = \frac{1}{(1+\Delta^2)}(uv^* + vu^*),$$

(3)

Coefficients $D_{ij}$ can be found as follows:

$$D_{xx} = \frac{\alpha_1(1-i\Delta)}{1+\Delta^2}(f_1 + f_2) - \alpha_2 + \ln\left(1 - \frac{\alpha_0}{1+\alpha_s(|u|^2+v^2)}\right), D_{yy} = \frac{\alpha_1(1-i\Delta)}{1+\Delta^2}(f_1 - f_2) - \alpha_2 +$$
$$\ln\left(1 - \frac{\alpha_0}{1+\alpha_s(|u|^2+v^2)}\right), D_{xy} = D_{yx} = \frac{\alpha_1(1-i\Delta)}{1+\Delta^2}f_3. \quad (4)$$

where:

$$f_1 = \left(\chi\frac{n_0}{2} - 1\right), \quad f_2 = \chi\frac{n_{12}}{2}, \quad f_3 = \chi\frac{n_{22}}{2}. \quad (5).$$

Here time t is normalized to the photon lifetime in the cavity $\tau_p$, $I_x=|u|^2$, $I_y=|v|^2$ are normalized to the saturation power $I_{ss}$ and $I_p$ is normalized to the saturation power $I_{ps}$, $\alpha_1$ is the total absorption of erbium ions at the lasing wavelength; $\alpha_2$ represents the normalized losses; $\alpha_0$ and $\alpha_s$ are parameters describing CNT saturable absorber; $\delta$ is the ellipticity of the pump wave, $\varepsilon=\tau_p/\tau_{Er}$ is the ratio of the the photon lifetime in the cavity $\tau_p$ to the lifetime of erbium ions at the first excited level $\tau_{Er}$; $\chi_{p,s}=(\sigma_a^{(s,p)}+\sigma_e^{(s,p)})/\sigma_a^{(s,p)}$, $(\sigma_a^{(s,p)}$ and $\sigma_e^{(s,p)}$ are absorption and emission cross-sections at the lasing (s) and pump (p) wavelengths); $\Delta$ is the detuning of the lasing wavelength with respect to the maximum of the gain spectrum (normalized to the gain spectral width), $\beta$ is the birefringence strength (2 $\beta=2\pi L_c/L_b$, $L_b$ is the beat length and $L_c$ is the cavity length),  We use an approximation in Eqs. (3) that the dipole moments of the absorption and emission transitions for erbium-doped silica are located in the plane orthogonal to the direction of the light propagation. This approximation results in the angular distribution of the excited ions n($\theta$), which can be expanded into a Fourier series as follows [11, 26-29]:

$$n(\theta) = \frac{n_0}{2} + \sum_{k=1}^{\infty} n_{1k}\cos(k\theta) + \sum_{k=1}^{\infty} n_{2k}\sin(k\theta),$$
$$f_1 = \left(\chi\frac{n_0}{2} - 1\right) + \chi\frac{n_{12}}{2}, \quad f_2 = \left(\chi\frac{n_0}{2} - 1\right) - \chi\frac{n_{12}}{2}, \quad f_3 = \chi\frac{n_{22}}{2}. \quad (6)$$

In contrast to more general approximation with 3D orientation distribution of the dipole orientations [31, 32], Eqs. (6) results in the finite dimension system presented by Eqs. (3) where only $n_0$, $n_{12}$, and $n_{21}$ components contribute to the laser dynamics. Unlike the other vector models, the vector model is in a simpler form presented by Eqs. (3) was justified in our previous publications on the mode-locked laser polarization dynamics at the time scales from one round trip to thousands of round trips [11, 26-29].

To obtain results shown in Figs. 4, we used the following parameters: a), d), g) $I_p$ =25, a=0.001 ; b), e), h) $I_p$ =22, a=0.02; c), f), i) $I_p$ =30, a=10. The other parameters: a)-i) $\beta_L = \beta_C = 0$, $\alpha_1 = 5.38$, $\alpha_2 = 1$, $\alpha_s = 10^{-3}$, $\alpha_0 = 0.136$, $\delta = 0.99$ (elliptically polarized pump SOP), $\Delta=0.13$ $\varepsilon=10^{-4}$, $\chi_p=1$, $\chi_s=2.3$, $\Omega=0.005\pi$, $\phi_0=0$, $\Psi=\pi/2$. To model the effect of the output POC transformation caused by the patchcord connect to the polarimeter, we use the following 3D rotation (around axes related to the Stokes parameters $S_0$, $S_1$, $S_2$, $S_3$) matrix [33]:

$$\begin{pmatrix}\tilde{S}_1\\\tilde{S}_2\\\tilde{S}_3\\\tilde{S}_0\end{pmatrix} = \begin{bmatrix}a_{11} & a_{12} & a_{13} & 0\\a_{21} & a_{22} & a_{23} & 0\\a_{31} & a_{32} & a_{33} & 0\\0 & 0 & 0 & 1\end{bmatrix}\begin{pmatrix}S_1\\S_2\\S_3\\S_0\end{pmatrix},$$

$$a_{11} = \cos(\psi)\cos(\gamma), \quad a_{12} = \cos(\gamma)\sin(\alpha)\sin(\psi) - \cos(\alpha)\sin(\gamma),$$

$$a_{13} = cos(\alpha)\,cos(\gamma)\,sin(\psi) + sin(\alpha)\,sin(\gamma),$$
$$a_{21} = cos(\psi)\,sin(\gamma),\ a_{22} = cos(\alpha)\,cos(\gamma) + sin(\alpha)\,sin(\psi)\,sin(\gamma),$$
$$a_{23} = -cos(\gamma)\,sin(\alpha) + sin(\psi)\,sin(\gamma),\ a_{31} = -sin(\gamma),\ a_{32} = cos(\psi)\,sin(\alpha),$$
$$a_{33} = cos(\alpha)\,cos(\psi). \tag{7}$$

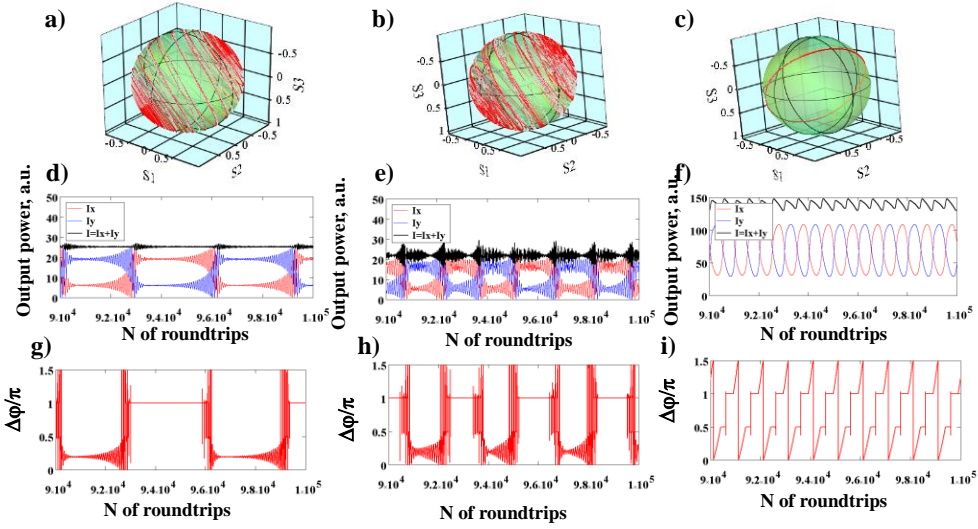

Fig. 4 Slow polarization dynamics: a)- c) trajectories on the Poincaré sphere.); d) - f) The output power vs number of the round trips for two linearly cross-polarized SOPs $I_x$ (blue line) and $I_y$ (red) and total power $I=I_x+I_y$ (black); ); g-f) the phase difference (red) vs number of the round trips. Parameters: $\alpha = -\pi/4$, $\beta = \pi/4$, $\gamma = 2\pi/5$, (a-i); $I_p$ =25 (a, d, g), $I_p$ =22 (b, e, h); Ip=30 (c, f, i). The time span of $10^4$ corresponds to 1 ms.

As follows from Fig. 4 (a-h), the injected signal with oscillating SOP modifies the spiral attractor and polarization dynamics of the polarization components x and y. The obtained theoretical results are quite close to the experimental data shown in Figs. 3 (a-h) that corresponds to the development of the soliton rain's condensed phase in Fig. 2 (a, b). With increasing the amplitude of the injected signal from a=0.02 to a=10, the polarization dynamics takes a completely different form, as shown in Fig. 4 (c, f, i). The spiral attractor is transformed to the circle (Fig. 4 c)), and x and y components' oscillations take the form of antiphase close to harmonic oscillations with the faster phase difference switching (Fig. 4 (f, i)). The dynamics is quite close to the experimental results shown in Fig.3(c, f, i) and the corresponding emergence of the soliton rain bunch shown in Fig. 2 (c).

Our previous publication shows that the mechanism driving the soliton rain origin and merging to the condensate phase is competition between polarization hole burning (PHB) caused by soliton rain pulses and holes refilling by the pump wave and active medium [19]. Here we are taking the next step by concluding the effect of the soliton rain on the active medium in modifying the polarization properties. In the soliton condensed phase (Fig. 2 (a, b, d, e)), the soliton rain depletes population inversion almost completely and cw component can't appears. The PHB is slightly modifying the active medium by inducing a small circular birefringence leading to the emergence of the spiral attractor on the Poincaré sphere (Fig. 3 (a, b, d, e, g, h) and Fig. 4 (b, e, h)). The soliton bunch emerges when PHB can't deplete the population inversion, and CW components appears (Fig. 2 (f)). The soliton bunch has a large amplitude and the periodically evolving SOP caused by the bunch drift. As a result, the spiral attractor is transformed to the circle on the Poincaré sphere (Fig. 3 (c, f, i) and 4 (c, f, i) ).

## 5. CONCLUSION
In conclusion, for the Er-doped fiber laser mode-locked using carbon nanotubes, we have demonstrated experimentally and theoretically that the soliton rain emergence can transform the polarization attractor on Poincaré sphere from spiral to the circle. Such experimentally observed transformation was explained based on the vector model developed by Sergeyev and co-workers in [11, 26-29] complemented by the vector SR model in terms of the injected signal with evolving SOP. The evolving SOP mimics the dynamically evolving polarization hole burning in the orientation distribution of Er ions' inversion caused by the main pulses and bunch of drifting soliton rain. The obtained results can pave the wave to development of a new technique for control of the multisoliton supramolecular structure in the form of the soliton rain that of interest for different application including spectroscopy, metrology and biomedical diagnostics.

**Funding.** Horizon 2020 ETN MEFISTA (861152) and EID MOCCA (814147).

**Disclosures.** The authors declare no conflicts of interest.